\documentclass[12pt]{article}

 \usepackage{mathrsfs}
   \usepackage{extpfeil}
 
\usepackage{epstopdf}
 \usepackage{cite}
\numberwithin{equation}{section}
\begin{document}

\begin{center}
{\LARGE{\bf{     $(1+1)$ Newton--Hooke Group for the Simple and Damped Harmonic Oscillator  }}}
\end{center}

\bigskip\bigskip

\begin{center}
 Przemys\l aw Brzykcy\footnote{E-mail address:  800289@edu.p.lodz.pl}
 \end{center}

\begin{center}

{\sl  Institute of Physics, Lodz University of Technology,\\ W\'{o}lcza\'{n}ska 219, 90-924 \L\'{o}d\'{z}, Poland.}\\
\medskip

\end{center}

\vskip 1.5cm
\centerline{\today}
\vskip 1.5cm

\begin{abstract}
It is demonstrated that, in the framework of the orbit method, a simple and damped
harmonic oscillators are indistinguishable at the level of an abstract Lie algebra.
This opens a possibility for treating the dissipative systems within the orbit 
method.
In depth analysis of the coadjoint orbits of the $(1+1)$ dimensional
Newton-Hooke group are presented.
Further, it is argued that the physical interpretation
is carried by a specific realisation of the Lie algebra of smooth functions on a
phase space rather than by an abstract Lie algebra.
\end{abstract}

%PACS numbers: 03.65.Ca, 42.50.-p

 \section{Introduction} \label{sec1}

Sidney Coleman famously said ``The career of a young theoretical physicist consists of treating the
harmonic oscillator in ever-increasing levels of abstraction''. 
The accuracy of this dictum is striking when one considers the abundance of scientific papers devoted 
to this subject across many branches of physics.
It has long been known that in the framework of the orbit method 
 \cite{kirillov1976elements,kirillov1999merits,kirillov2004lectures}
the oscillator is described  by the  Newton--Hooke  ($NH$) group.
The $NH$  type groups first appeared in the classification of the possible kinematical 
groups  \cite{Bacry}. 
The thorough study of $(3+1)$ dimensional $NH$ group was presented in \cite{Derome}.
The orbit method was employed in   \cite{NH3} to study the centrally extended $NH$ group in $(2+1)$ 
dimensions. The coadjoint orbits and the irreducible  representations were calculated therein.
Besides the orbit method  a planar system with exotic Newton--Hooke symmetry was
constructed by the technique of nonlinear realisation \cite{Alvarez20071556}, the analysis therein 
included the chiral decomposition.
 The idea of chiral decomposition was later applied to the non-commutative Landau problem \cite{NH2,Zhang}
and  to the  rotation-less $NH$ symmetry of the $3$D anisotropic oscillator 
\cite{NH5}.
Some work on the anisotropic (2+1) dimensional  $NH$ group was   presented in  \cite{NH4}.
The extended conformal $NH$ type symmetries were also studied  in connection to the Pais--Uhlenbeck oscillator \cite{PU1,PU,NH6}.
This short overview is far from being complete but it shows, how telling a study of this simple system can be.
It is worth to mention that orbit method was also successfully used to analyse systems with
Galilei and Poincar\'e type symmetries both in the free case and with external electromagnetic
fields \cite{lukierski1997galilean,stichel,anyons,Jackiw2000237,Ghosh2006350,Duval2000284,0305-4470-34-47-314,Duval:2001hu,1126-6708-2002-06-033,delOlmo20062830,SIGMA}.

In this paper an accessible yet illuminating example of   harmonic oscillator  is examined in the framework
of the orbit method  \cite{kirillov1976elements,kirillov1999merits,kirillov2004lectures}.
In the case of simple harmonic oscillator the Lie algebra of $(1+1)$ $NH$ group is derived from
the standard Hamiltonian description by a technique encouraged by \cite{stichel}.
Detailed analysis of the coadjoint action provides a full understanding of the physical interpretation. 
Clearly, analysis becomes more involved  for the  dissipative systems. 
However,  a proper canonical transformation may  allow  for a significant   simplification. 
For example, the damped harmonic oscillator  can be described by the same Lie algebra as   the undamped case.
This simplification comes at a price of using rather elaborate coordinates.
Consequently, at the level of the Lie algebra the damped harmonic oscillator is indistinguishable from the 
undamped one. 
This example illustrates a possible way of treating the dissipative systems within the 
framework of the orbit method.
Apparently an abstract Lie algebra does not carry the physical interpretation of the system.  
The question arrises how to use the orbit method  so that the physical  interpretation is not lost. 
 The current paper is devoted to just this investigation.

This paper is structured as follows.
In Section \ref{NH}, starting with the Hamiltonian of the harmonic oscillator,
the Lie algebra of the $(1+1)$ dimensional Newton--Hooke group is derived to set the scene for the further analysis.
Section \ref{coadjoin} provides the coadjoint action of the group under
investigation. Also, the symplectic structure on the coadjoint orbit of $(1+1)$ Newton--Hooke
group are given. The in depth analysis of the coadjoint orbits is presented in
Section \ref{orbits}.
Section \ref{damped} is devoted to the damped harmonic oscillator and shows that
it may be described by the same abstract Lie algebra as the undamped case.
The paper closes with   conclusions in Section \ref{conclusions}
where also some outlooks are provided.

  \section{The (1+1) Newton-Hooke group}\label{NH}

To focus the attention take the Hamiltonian of the simple harmonic oscillator
  \begin{equation}
    h(p,x)= \frac{p^2}{2m}+ \frac{m\omega^2 x^2}{2}, \label{harmonic_hamiltonian}
  \end{equation}
  where $p$ is the kinematic momentum and   $x$ is the displacement of the
  oscillator.   Exploiting the canonical Poisson bracket
  \begin{equation}
    \{ F(p,x),  G(p,x)\} = \frac{\partial F}{\partial x}  \frac{\partial G}{\partial p}
    - \frac{\partial F}{\partial p} \frac{\partial G}{\partial x} \label{harmonic_poisson_1}
  \end{equation}
one arrives at the well known equations of motion
  \begin{equation}
      \dot{p}= -m\omega^2 x,\quad
      \dot{x}=   \frac{p}{m}
\label{harminic_motion_1}
  \end{equation}
which when put together read  $\ddot{x}=-\omega^2 x$.
In order to use the orbit method to describe the simple harmonic oscillator an
appropriate Lie algebra is needed. This algebra should be such that the
equations of motion on its coadjoint orbits are equivalent to  (\ref{harminic_motion_1}).
Herein such a Lie algebra is constructed starting with the algebra of smooth functions on the phase space equipped  with the Poisson bracket
(\ref{harmonic_poisson_1}).

The method of constructing such a Lie algebra is based on  Poisson's theorem stating that the Poisson bracket of two quantities that
are constants of motion is also a constant of motion.
The Hamiltonian (\ref{harmonic_hamiltonian}) i.e. the total energy of the system is the only integral of motion.
This system also admits    constants of motion e.g $f(p,x,t)= t-\frac{1}{\omega} \arctan{\frac{\omega k}{p}}$
which at  $p=0$ has to be understood in the sense of the limit.
  It is mentioned here for the sake of completeness, however will not be
  utilised in the present paper because there is no need to consider the time dependent generators.
  Therefore,  $h$ should be included in the set of generators.
    Inasmuch as some coordinates are needed, one just checks
  whether $p$ and $x$ could do  the job. To this end  calculate the Poisson bracket (\ref{harmonic_poisson_1})
  for all the pairs selected from the set $\{h,p,x\}$ and find that the    non-vanishing brackets are
   \begin{equation}
       \{ h, p  \}= m \omega^2 x,\quad
       \{ h, x \}=     -\frac{1}{m} p,\quad
       \{ x, p  \} = 1.
  \end{equation}
Quick conclusion is that, in order to have a closed algebra,  the $\{h,p,x\}$ ought to be augmented by a constant
  function equal to 1.
  Even more elegantly one may replace $x$ with $k=mx$ and use a constant function equal
  $m$. In which case the Hamiltonian (\ref{harmonic_hamiltonian})
  becomes
   \begin{equation}
    h(p,k)= \frac{p^2}{2m}+ \frac{\omega^2 k^2}{2m}, \label{harmonic_hamiltonian2}
  \end{equation}
  and the Poisson bracket (\ref{harmonic_poisson_1}), by the chain rule, reads now
  \begin{equation}
    \{ F(p,k),  G(p,k)\} =  m \left( \frac{\partial F}{\partial k}  \frac{\partial G}{\partial p}
    - \frac{\partial F}{\partial p} \frac{\partial G}{\partial k} \right).\label{harmonic_poisson_2}
  \end{equation}
The non-vanishing Poisson brackets are
      \begin{equation}
       \{ h, p  \}=  \omega^2 k,\quad
       \{ h, k \}=     - p,\quad
       \{ k, p  \} = m.
  \end{equation}
Therefore the functions $J_1=m, J_2=h(p,k),  J_3=p, J_4=k$  span the Poisson algebra
under   the Poisson bracket (\ref{harmonic_poisson_2}).
  Note that $J_1$ is a central generator.
What was described above is known as the Lie algebra of the $(1+1)$ dimensional
Newton-Hooke group. At the abstract level it is a four dimensional Lie algebra
spanned by $J_1=M,J_2=H,J_3=P,J_4=K$ characterised by the following
  nonzero structure constants
  ${c_{23}}^4=\omega^2$,  ${c_{24}}^3=-1$, ${c_{34}}^1=-1$, in the above numbering of the basis, which will be kept throughout this paper.

\section{Coadjoint action and dynamics}\label{coadjoin}

The matrices of the adjoint action $m^{ad}_{J_i}$ corresponding to the generators $J_1,\dots, J_4$ are given by
$(m^{ad}_{J_i})_{jk} = {c_{ik}}^j$ where ${c_{ik}}^j$ are the structure constants.
Explicitly  $$  \mathrm{m}^{Ad}_{J_2}=      \begin{bmatrix}
 0 & 0 & 0 & 0  \\
 0 & 0 & 0 & 0  \\
   0 & 0 & 0 & -1  \\
     0 & 0 & \omega^2 & 0
\end{bmatrix},
\mathrm{m}^{Ad}_{J_3}= \begin{bmatrix}
 0 & 0 & 0 & -1  \\
 0 & 0 & 0 & 0  \\
   0 & 0 & 0 & 0  \\
     0 & -\omega^2 & 0 & 0
\end{bmatrix},
     \mathrm{m}^{Ad}_{J_4}= \begin{bmatrix}
 0 & 0 & 1 & 0  \\
 0 & 0 & 0 & 0  \\
   0 & 1 & 0 & 0  \\
     0 & 0 & 0 & 0
\end{bmatrix}
$$
and  $\mathrm{m}^{Ad}_{J_1}$ is the zero matrix since $m$ is a central  generator.
  The generic element $g$ of our group can be written as
 $ g = (\eta,b,a,v)= e^{\eta M} e^{bH} e^{aP}e^{vK}\in G$.
The coordinates for the dual to our Lie algebra $\mathfrak{g}^*$ are realised by $m,h,p,k$.
The matrix of the coadjoint
  action of an element $g\in G$ is then given by $M^{coAd}_g=e^{ -v \, \mathrm{m}^{Ad}_{K} }
  e^{ -a \, \mathrm{m}^{Ad}_{P} }
  e^{ -\tau \, \mathrm{m}^{Ad}_{H} }
  e^{ -\eta \, \mathrm{m}^{Ad}_{M} }
  $
  which explicitly reads
\begin{equation}
     M^{coAd}_g= \begin{bmatrix}
 1  & \frac{v^2+a^2\omega^2}{2} & -v \cos{b\omega}-a \omega \sin{b\omega}  & a \cos{b\omega}-\frac{v}{\omega}  \sin{b\omega}  \\
 0 & 1                                                 &  0                                                                           & 0  \\
 0 &  -v                                              & \cos{b\omega}                                                    &\frac{\sin{b\omega}}{\omega}  \\
 0 & a\omega^2                               &-\omega \sin{b\omega}                                     & \cos{b\omega}
\end{bmatrix}.
\end{equation}
An element of $\mathfrak{g}^*$ is represented as a row vector $ \xi=[ m,h,p,k]$.
Then  the coadjoint action of $g\in G$ is calculated by matrix multiplication of
$\xi$ by $\mathfrak{g}^*$ on the right, which yields the following explicit form of the coadjoint action
\begin{equation}\left\{
 \begin{array}{l l}
   m'&=m,\\
      h'&=h+ \frac{1}{2}m v^2 +\frac{ma^2\omega^2}{2}-vp+a\omega^2 k,\\
        p'&=(p-mv)  \cos{b\omega}-\omega (ma+k) \sin{b \omega},  \\
            k'&=(ma+k)\cos{b\omega} +\frac{p-mv}{\omega} \sin{b \omega}.
 \end{array} \right.\label{coad_action}
 \end{equation}
 More detailed  analysis of the action (\ref{coad_action}) will be presented in
 the following section.
The  next step is to  calculate the invariants of the coadjoint action i.e. smooth
 functions $C$ on $\mathfrak{g}^*$, such that
$
 \forall_{g\in G}\, \forall_{\xi\in \mathfrak{g}^*}\quad
 C(\mathrm{coAd}_g(\xi))=C(\xi).
$
They are solutions to the
following set of differential equations \cite{Beltrametti196662,doi:10.1063/1.522727,doi:10.1063/1.522992,0305-4470-39-20-009,snobl2014classification}
   \begin{equation}
      \begin{bmatrix}
0 & 0 & 0& 0 \\[0.2cm]
 0 & 0 & \omega^2 k  & -p \\[0.2cm]
   0 & -\omega^2 k & 0 &-m  \\[0.2cm]
     0 & p &m &0  \\[0.2cm]
\end{bmatrix}
   \begin{bmatrix}
\frac{\partial C}{\partial m}\\[0.2cm]
\frac{\partial C}{\partial h}  \\[0.2cm]
\frac{\partial C}{\partial p}  \\[0.2cm]
\frac{\partial C}{\partial k}  \\[0.2cm]
\end{bmatrix}=
\begin{bmatrix}
0\\[0.2cm]
0\\[0.2cm]
0\\[0.2cm]
0\\[0.2cm]
\end{bmatrix}.\label{harmonic_inv_set}
\end{equation}
There are two solutions to (\ref{harmonic_inv_set}), namely
\begin{equation}
C_1=m,\quad
C_2=k^2 - \frac{2mh}{\omega^2}+ \frac{p^2}{\omega^2}.
\end{equation}
The first one is a trivial  consequence of $m$ being a central generator.
Consider a map
\begin{equation}
 \begin{split}
   C:\mathfrak{g}^* &\to \mathbb{R}^2,\\
   \xi  &\mapsto\left(C_1(\xi),C_2(\xi)\right).
 \end{split}\label{harmonic_inv_map}
\end{equation}
At each point $\xi' = coAd_g(\xi)$, $g\in G$ of the orbit through $\xi$ the value of (\ref{harmonic_inv_map})
is constant.
Moreover, mapping (\ref{harmonic_inv_map}) is of a constant and maximal rank, therefore the preimage of a point is a
submanifold in $\mathfrak{g}^*$. Each of its compact components  is precisely a coadjoint orbit through
$\xi$.
In the present case, the orbit, denoted $\mathcal{O}^{C_1,C_2}$, admits a single global parametrisation
\begin{equation}
\varphi: (p,k)\mapsto \left(m=C_1, h=\tilde{h}=\frac{p^2}{2m}+\frac{\omega^2 k^2}{2m} - \frac{\omega^2 C_2}{2m} ,p,k\right)
\end{equation}
so, in principle, it might be covered by a single map e.g. $\varphi^{-1}$.
Note that, in the this example,  the hamiltonian
\begin{equation}
 \tilde{h}(p,k)=\frac{p^2}{2m}+\frac{\omega^2 k^2}{2m} - \frac{\omega^2 C_2}{2m} \label{harmonic_ham_inv}
\end{equation}
which was derived from the invariants of the coadjoint action is, up to an
additive constant, equivalent to   the initial Hamiltonian
(\ref{harmonic_hamiltonian}).
For the sake of completeness note that, the Jacobian  of the map
 $
 (m,h,p,k)\mapsto \left(C_1, C_2 ,p,k\right)
$
is $-\frac{2m}{\omega^2}$ so there is a singularity for $m=0$.
This case shall not be discussed  in depth for it is of no physical interest.
It suffices to say that, for the fixed $m=0$, there is one invariant of the
coadjoint action
$
 C = k^2 + \frac{p^2}{\omega^2},
$
and since $h$ is unrestricted, the orbits resemble the flatten  cylinders.
In what follows,   $m\neq 0$ shall be assumed.
The Poisson tensor $\Lambda$ on the orbit  $\mathcal{O}^{C_1,C_2}$, written in the chart
  $(\mathcal{O}^{C_1,C_2},\varphi^{-1} )$ reads
  \begin{equation}
   \Lambda^{ij}=      \begin{bmatrix} 0 & -m   \\  m & 0 \end{bmatrix},
  \end{equation}
  which is equivalent to (\ref{harmonic_poisson_2}).
Since $m\neq 0$ on the orbit $\mathcal{O}^{C_1, C_2}$  the  Poisson structure is non-degenerate and
one quickly finds, by the techniques presented in \cite{fortschr}, that the corresponding symplectic two-form is
\begin{equation}
 \omega= -\frac{1}{m} \mathrm{d}p \wedge \mathrm{d} k.
\end{equation}
Therefore, employing the Hamiltonian (\ref{harmonic_ham_inv}),  one finds the
equations of motion to be
\begin{equation}
 \dot{p}= - \omega^2 k, \quad \dot{k}=p\label{harmonic_motion_2}
\end{equation}
  which, when combined with $k=mx$, are equivalent to (\ref{harminic_motion_1}).

\section{Coadjoint orbits}\label{orbits}
Further insight into the structure of the coadjoint orbit can be gained by
examining, one by one, the coadjoint actions of the group elements that
correspond to the generators.
Consider a test solution of (\ref{harmonic_motion_2}) for example $p=-m\omega A \sin{(\omega t)},k=mA\cos{(\omega t)}$
   that is, at $t=0$ the displacement is maximal and momentum is zero.
The energy is constant and at any time is given by (\ref{harmonic_ham_inv}) ($E=\tilde{h}(p,k)$), furthermore $m=C_1$ is also fixed.
The trajectory of the system, as time flies, is then given by
 \begin{equation}
\xi(t)=[m,E=\tilde{h}(p,k),p=-m\omega A \sin{\omega t},k=mA\cos{\omega t}].
\label{harmonic_trajectory}
\end{equation}
The trajectory (\ref{harmonic_trajectory}) lies on the coadjoint orbit
characterised by $C_1=m$ and a fixed $C_2$.

Coadjoint action of  a group element     $g=\exp{(\tau H)}$
generated by $H$  on (\ref{harmonic_trajectory}) is: $m' = m$, $E' = E$ and
   \begin{equation}\nonumber
     \begin{split}
       p'&=-m\omega A \sin(\omega (t+\tau)),\\
       k'&=mA \cos{(\omega (t+\tau))}
     \end{split}
   \end{equation}
that is to say $m$ and $E$ remain constant and $p$, $k$ follow the elliptic
trajectory. Clearly, $H$ generates temporal shifts.
Moreover as the system evolves, it stays on the same orbit.

Next, let us consider a group element     $g=\exp{(l P)}$
generated by $P$.   Its coadjoint action on (\ref{harmonic_trajectory}) is: $m' =m$, $p'=p$
and
     \begin{equation}\nonumber
     \begin{split}
       E' &= E + \frac{1}{2} m \omega^2 x'^2 + mx' \omega^2 A \cos{\omega t},\\
       k'&=k+ml.
     \end{split}
   \end{equation}
The displacement ($x=\frac{k}{m}$) is increased by $l$ and energy is changed exactly in such a way, that the system stays on the
 same orbit which means that $P$ generates spatial shifts of the initial  conditions.

  \begin{figure}[h!]
 \centering
  \includegraphics[width=1\textwidth]{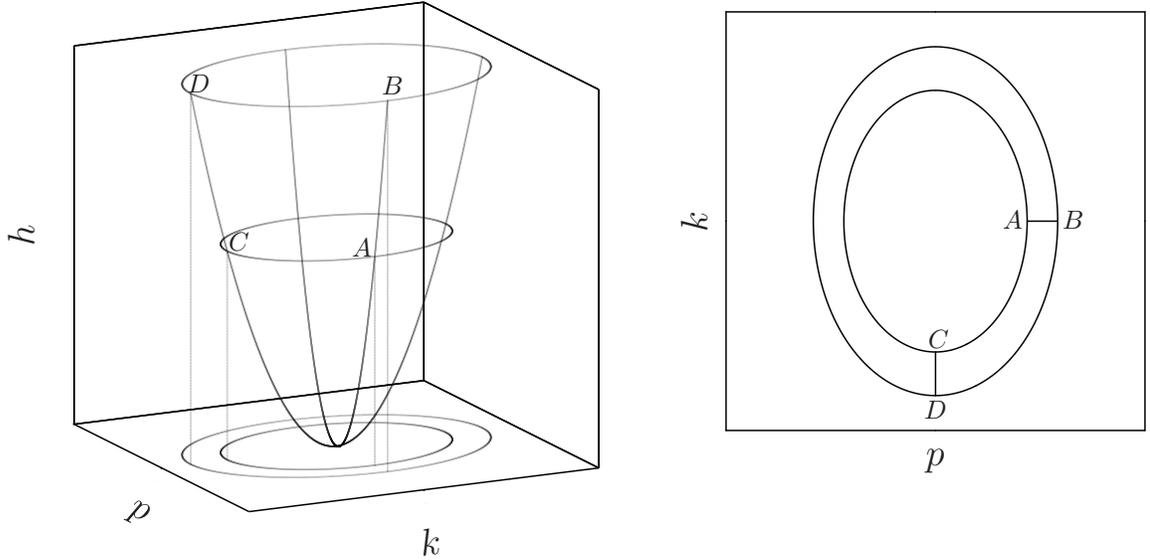}
 \caption{Ilustration  of the coadjoint action of the elements of the group $NH(1+1)$ corresponding to the generators.
 The time evolution brings the system along the elliptical trajectory e.g. from
 point $A$ to $C$ or $B$ to $D$.
 Starting from the point $A$ the system can be moved to point $B$ by the
 spatial shift generated by $P$.
 Performing a boost generated by $K$
brings the system from the point $C$ to $D$.
 }
 \label{fig:harmonic_1}
\end{figure}

Finally, a group element     $g=\exp{(u K)}$generated by $K$ acts on (\ref{harmonic_trajectory})
as: $  m' = m$, $k'=k$
         \begin{equation}\nonumber
     \begin{split}
       E' &= E + \frac{1}{2} mu^2 +mu \omega A \sin{\omega t},\\
       p'&=p-mu
     \end{split}
   \end{equation}
 i.e. $m$ and $k$ are constant, momentum $p$ is decreased by $mu$ and energy
 is adjusted so that the system remains on the orbit.
Clearly, $K$ generates the momentum shifts.
For the sake of completeness it is worth mentioning  that the action
of group elements generated by $M$ is identity.
The examples of above-described actions are presented in Figure \ref{fig:harmonic_1} which
also, by a small leap of imagination, allows us to visualise the coadjoint orbit.

\section{Damped harmonic oscillator} \label{damped}
A slightly more complex system which  can be investigated by similar techniques is a damped harmonic oscillator.
Take the following time dependent Hamiltonian
     \begin{equation}
    h(\mathcal{P},q,t) = \frac{ \mathcal{P}^2}{2m} e^{-2\gamma t} + \frac{1}{2}m \omega_0^2 e^{2\gamma t}
    x^2\label{damped_hamiltonian}
  \end{equation}
  where
  $2\gamma = \frac{\beta}{m}$ with $\beta$ being the friction coefficient and  $\omega_0$ is the undamped frequency of the oscillator.
  Note that $\mathcal{P} = m \dot{x} e^{2\gamma t}$ i.e. the canonical momentum does not coincide with the 
kinetic momentum $p=m\dot{x}$.
  The Hamiltonian (\ref{damped_hamiltonian})   with the canonical  Poisson bracket
 yields the following equations of motion
     \begin{equation}
       \dot{x}  = \frac{\mathcal{P}}{m} e^{-2 \gamma t},\quad
      \dot{\mathcal{P}}  = -m\omega_0^2 e^{2\gamma t}x
   \end{equation}
or equivalently $     \ddot{x}+2\gamma \dot{x} + \omega_0^2x=0.$
 It is an easy exercise to check that the procedure
that was carried out in  Section \ref{NH} for the undamped
  oscillator fails in the present case.
Indeed, introducing the new coordinate $k=mx$ one finds that the hamiltonian (\ref{damped_hamiltonian})
becomes
\begin{equation}
    h(\mathcal{P},k,t) = \frac{\mathcal{P}^2}{2m} e^{-2\gamma t} + \frac{\omega_0^2   k^2}{2m}  e^{2\gamma t}
\end{equation}
  and, by the chain rule, new the Poisson bracket is just
  (\ref{harmonic_poisson_2}).
Then, one quickly finds that
\begin{equation}
\left\{ h,\mathcal{P}\right\}=\omega_0^2 e^{2\gamma t} k,\quad
\left\{ h,k\right\}=-e^{-2\gamma t}
\mathcal{P} ,\quad \left\{ k, \mathcal{P} \right\}=m
\end{equation}
which fails to constitute a  Lie algebra because there is an undesired time
dependency of the structure constants.
One way to deal with this problem  is to use a generating function method to bring the Hamiltonian (\ref{damped_hamiltonian})
to a more convenient form (see e.g. \cite{greiner2009classical}).
Consider the following generating function of  the second kind
\begin{equation}
  F_2(k,P,t) = e^ {\gamma t} k P - \frac{1}{2} m \gamma e^{2\gamma t} x^2.
\end{equation}
  The transformation rules for the coordinates are
  \begin{equation}
      \mathcal{P}=\frac{\partial F_2}{\partial x} =  e^{\gamma t}P - m \gamma e^{2\gamma t}
      x,\quad
      Q=\frac{\partial F_2}{\partial P} = e^{\gamma t} x\\
  \end{equation}
furthermore,   the old $h$ and the new $H$ Hamiltonians obey
\begin{equation}
        H-h = \frac{\partial F_2}{\partial t} = \gamma e^ {\gamma t} x P -m \gamma^2 e^{2\gamma t}
        x^2=\gamma QP - m \gamma^2 Q^2. \label{damped_transf}
\end{equation}
  The relations between old $(\mathcal{P},x)$ and new $(P,Q)$ coordinates can be  written     as
  \begin{equation}
    \begin{bmatrix} \mathcal{P} \cr x  \end{bmatrix} =      \begin{bmatrix} e^{\gamma t}  & -m \gamma e^{\gamma t} \cr 0 & e^{-\gamma t} \end{bmatrix}     \begin{bmatrix} P \cr Q
    \end{bmatrix},
      \quad
           \begin{bmatrix} P \cr Q \end{bmatrix} =      \begin{bmatrix} e^{-\gamma t}  & m \gamma e^{\gamma t}\cr 0& e^{\gamma t} \end{bmatrix}     \begin{bmatrix} \mathcal{P} \cr x
             \end{bmatrix}.
  \end{equation}
By the chain rule
  $\frac{\partial{ }}{\partial{\mathcal{P}}} =  \frac{\partial{P}}{\partial{\mathcal{P}}}  \frac{\partial{ }}{\partial{P}}+  \frac{\partial{Q}}{\partial{\mathcal{P}}}
 \frac{\partial{ }}{\partial{Q}}= e^{-\gamma t}
 \frac{\partial{ }}{\partial{P}}$ and
 $  \frac{\partial{ }}{\partial{x}} =  \frac{\partial{P}}{\partial{x}}  \frac{\partial{ }}{\partial{P}}+  \frac{\partial{Q}}{\partial{x}}
 \frac{\partial{ }}{\partial{Q}} = m\gamma e^{\gamma t}    \frac{\partial{ }}{\partial{P}}
 + e^{\gamma t}   \frac{\partial{ }}{\partial{Q}}$
so one quickly finds that the Poisson bracket (\ref{harmonic_poisson_1}) becomes
 \begin{equation}
    \{ F(P,Q),  G(P,Q)\} = \frac{\partial F}{\partial Q}  \frac{\partial G}{\partial P}
    - \frac{\partial F}{\partial P} \frac{\partial G}{\partial Q}
  \end{equation}
  i.e. the transformation (\ref{damped_transf}) is canonical but, since $H\neq h$ it is not a
  symmetry.
Finally, the transformed Hamiltonian takes the following form
\begin{equation}
 H(P,Q)=\frac{P^2}{2m} + \frac{1}{2} m (\omega_0^2-\gamma^2)Q^2\label{damped_H2}
\end{equation}
  which, functionally is just (\ref{harmonic_hamiltonian}) with
  $\omega^2=\omega_0^2-\gamma^2$.
Therefore, in the new coordinates $P$ and $Q$   the procedure of
constructing the Lie algebra as in Section \ref{NH} can be carried out.
The resulting algebra is exactly (1+1)
Newton-Hooke algebra as it was for the undamped oscillator
therefore, at the level of the abstract Lie algebra the two systems are
indistinguishable.
The difference lies in the realisation of the generators as smooth functions of the phase space coordinates.
It is important to stress that the Hamiltonian derived from the invariants of
the coadjoint action would be functionally equivalent (up to an additive constant)
to (\ref{damped_H2}) not to the initial Hamiltonian (\ref{damped_hamiltonian}) as the interpretation
of the coordinates has changed.

\section{Concluding remarks}\label{conclusions}
It was shown that, in the framework of the orbit method, a simple and damped
harmonic oscillators can be described by the same abstract Lie algebra.
 The simple, yet striking example presented here shows that, when the dynamics
on the coadjoint orbits  are considered, a simple knowledge of an abstract Lie
algebra does not suffice to provide the physical interpretation.
What describes the system is rather a specific realisation of the Lie algebra
in terms of the smooth functions on the classical phase space.

 Particularly, the result presented in the current paper
stresses the importance of keeping track of the physical interpretation when
constructing the dynamics on the coadjoint orbits which might be crucial when
the deformation quantisation on the coadjoint orbit is considered
\cite{Gadella,moyal1,moyal2,Martin,moyal3,moyal4,Dito2006309}.
One way to achieve that is to derive a relevant Lie algebra starting from the
Hamiltonian formulation as was done in the current paper for the harmonic
oscillator or in \cite{stichel} for extended Galilei group also known in the
literature as the Galilei-Maxwell group \cite{anyons}.
It is the intention of the author to follow with the application of the current
results in case of the Poincar\'e-Maxwell group soon.

\end{document}